\documentclass{epl}

\usepackage[centertags]{amsmath}
\usepackage{psfrag}
\usepackage{subfigure}

\newcommand{\lqq}{``}
\newcommand{\rqq}{''}
\newcommand{\osum}[1]{\operatorname*{\overline{\sum}}_{#1}}
\newcommand{\psum}[1]{\operatorname*{\sum^\prime\nolimits\!}_{#1}\,}
\newcommand{\olsum}[1]{\operatorname*{\overline{\sum}}_{#1}\limits}
\newcommand{\plsum}[1]{\operatorname*{\sum^\prime\nolimits\!}_{#1}\limits\,}
\newcommand{\chiprep}{\chi_{\rm p}}
\newcommand{\chimeas}{\chi_{\rm m}}
\newcommand{\chicrit}{\chi_{\textrm{crit}}}
\newcommand{\lenprep}{\xi_{\textrm{p}}}
\newcommand{\lenmeas}{\xi_{\textrm{m}}}
\newcommand{\lenloc}{\xi_{\textrm{l}}}
\newcommand{\scalf}{w}

\title{Glassy states and microphase separation in cross-linked homopolymer
  blends}

\shorttitle{ Glassy states in cross-linked polymer blends}

\author{ Christian Wald\inst{1}%
  \thanks{E-mail: \email{wald@theorie.physik.uni-goettingen.de}}
  \and Annette Zippelius\inst{1} \and Paul
  M.~Goldbart\inst{2}} \shortauthor{C. Wald \etal}

\institute{
\inst{1}Institut f\"ur Theoretische Physik -
  Georg-August-Universit\"at G\"ottingen,\\
  Friedrich-Hund-Platz~1, 37073 G\"ottingen, Germany\\
\inst{2}Department of Physics -
University of Illinois at Urbana-Champaign,\\
1110 West Green Street, Urbana, Illinois 61801-3080, U.S.A.}


\begin{document}

\maketitle

\begin{abstract}
  The physical properties of blends of distinct homopolymers, cross-linked
  beyond the gelation point, are addressed via a Landau approach involving a
  pair of coupled order-parameter fields: one describing vulcanisation, the
  other describing local phase separation.  Thermal concentration
  fluctuations, present at the time of cross-linking, are frozen in by
  cross-linking, and the structure of the resulting glassy fluctuations is
  analysed at the Gaussian level in various regimes, determined by the
  relative values of certain physical length-scales.  The enhancement, due to
  gelation, of the stability of the blend with respect to demixing is also
  analysed.  Beyond the corresponding stability limit, gelation prevents
  complete demixing, replacing it by microphase separation, which occurs up to
  a length-scale set by the rigidity of the network, as a simple variational
  scheme reveals.
\end{abstract}


\section{Introduction}

Molten blends of distinct homopolymers have a strong tendency to phase
separate, compared with unpolymerised mixtures, due to the fact that their
entropy of mixing is reduced by a factor of the degree of polymerisation.
Random permanent cross-linking hampers the demixing process, and completely
inhibits macroscopic phase separation, provided enough cross-links are
introduced to cause the blend to undergo a transition to a gel.
De~Gennes~\cite{deGennes1979} was the first to point out that the region of
compatibility ({\it i.e.\/}~the region in which the melt remains mixed) is
substantially increased by cross-linking, and that instead of macroscopic
phase separation, the gel undergoes microphase separation (MPS) with a domain
size comparable to the typical mesh size of the random macromolecular network.

De~Gennes' predictions were subsequently confirmed in scattering
experiments~\cite{Bauer1988}.  However, a discrepancy remained, concerning the
scattering intensity at small wave-numbers.  Whereas de~Gennes predicted a
{\it vanishing\/} intensity at zero wave-number, via an analogy with the
polarisation of a dielectric medium, the experiments showed a {\em
  non-vanishing\/} intensity.  Several modifications of the de~Gennes theory
have been proposed.  Benmouna {\it et al.\/}~\cite{Benmouna1994} suggest that
the discrepancy arises through the neglect of concentration fluctuations,
which are present during the cross-linking process and are \lqq frozen in\rqq\ 
by the cross-linking process.  These authors introduce a screening length,
self-consistently, by assuming that the scattering intensity at zero
wave-number is not changed by cross-linking.  Read {\it et
  al.\/}~\cite{Read1995} instead suggest a microscopic model for the gel, in
which each chain is anchored at its ends to fixed random points in space. By
using reasonable assumptions for the distribution of the quenched
end-to-end-vectors, they arrive at an intensity similar to that of
Ref.~\cite{Benmouna1994}, and also investigate effects of applied strain.
Simulations by Lay {\it et al.\/}~\cite{Sommer2000} reveal rather large
cooperative rearrangements of the network upon MPS, allowing for domain sizes
larger than the radius of gyration of the network strands.

In this Letter we focus on \lqq charge\rqq\ fluctuations in gels that
have been prepared from blends in a homogeneously mixed state,
characterised by a sufficiently small Flory incompatibility
parameter. By \lqq charge\rqq\ we mean the difference between the
local densities of the two species of homopolymer.  We concentrate on
two themes: \begin{itemize} \item{ To what extent do the charge
fluctuations that are present during cross-linking become frozen-in in
the gel phase?  This is particularly interesting if the cross-linking
is performed close to phase separation, so that fluctuations are
present up to a very large length-scale, which can be either smaller
or larger than the localisation length-scale characterising the gel.}
\item{How do frozen-in charge fluctuations affect the scattering
intensity and MPS?  To what extent does gelation enhance the stability
of the mixed state, and to what extent can one characterise the MPS
state that emerges at large enough incompatibility parameter?}
\end{itemize}

Our starting point is a Landau expansion for the free energy in
terms of {\it two\/} order parameter fields: the local charge
fluctuations associated with phase separation, and the local static
density fluctuations associated with the gelation transition.  This
free energy can be derived from a microscopic model~\cite{Wald2004},
extending previous work on the gelation transition~\cite{GoldbartAdvPhy1996}
to include the incompatibility of homopolymer blends.  In this Letter we
do not dwell upon the derivation of the Landau free energy, but
rather work out its consequences.  As we shall see, the phase diagram
is controlled by three parameters:
the cross-link density control parameter $\mu$, and
the incompatibilities
at cross-linking $\chiprep$ 
and at measurement $\chimeas$ (${\rm p}$ stands for {\it preparation},
and ${\rm m}$ for measurement).

The Landau free energy allows us to compute the frozen-in
charge fluctuations without any {\it ad hoc\/} assumptions.
We show that there are competing length-scales:
in a strong gel the fluctuations are frozen-in almost completely,
and hence preserve the native length-scale at cross-linking;
in a weak gel, by contrast, the charge fluctuations are only
partially frozen in, limited by the length-scale characterising localisation
in the gel. We calculate the modification of the range of compatibility 
due to cross-linking, in terms of the gel order parameter, and
discuss the scattering intensity in the gel, including the region
in which MPS is approached.  With the aim of developing a rough
picture of the MPS state itself, we also analyse a lamellar-state
Ansatz for the MPS state beyond, but close to, the transition.


\section{Model}

We consider a blend of two incompatible homopolymer species, \lqq{\it
  A\/}\rqq\  and \lqq{\it B\/}\rqq.  The two species are modelled identically
  as Gaussian chains of length $L$, and the blend is taken to contain $N/2$
  chains of each type. The mutual repulsion is modelled by the interaction  
\begin{equation}
  H^\chi = -\frac{\chi V}{2N}
  \sum_{a,a^\prime=A,B} (2\delta_{a,a^\prime}-1)
  \sum_{i,i^\prime=1}^{N/2}
  \int_0^1\!ds\int_0^1\!ds^\prime\
  \delta\big(\bm{r}_{i}^a(s)-\bm{r}^{a^\prime}_{i^\prime}(s^\prime)\big)\,,
\end{equation}
where $\bm{r}_i^a(s)$ denotes the position of the monomer at arclength
$s$ on chain $i$, which can be either of type {\it A\/} ($a=A$) or {\it B\/}
($a=B$). The strength of the repulsion is controlled by the incompatibility,
{\it i.e.\/}, the Flory-Huggins parameter $\chi$.  On cooling, the
un-cross-linked melt would undergo a second-order phase transition at
$T/\chi=1$ from a homogeneously mixed phase to a phase-separated state. In the
following, we choose units such that $k_{\textrm{B}} T=1$.  Since there is no
other relevant energy scale, we shall refer to $\chi$ as the inverse
temperature. The order parameter for a general phase-separation transition is
given by the local imbalance in the concentration of {\it A\/} and {\it B\/}
monomers (which 
we are referring to as charges): 
\begin{equation}
\Psi(\bm{k})=\frac{1}{N}
  \sum_{a=A,B}\sum_{i=1}^{N/2}\int_0^1\!ds\,
  (\delta_{a,A}-\delta_{a,B})\,
 \langle e^{i\bm{k}\cdot\bm{r}_i^a(s)}\rangle.
\end{equation}
In the homogeneous state the expectation value of the order parameter
vanishes: $\Psi(\bm{k})=0$. If two homogeneous phases coexist,
$\Psi(\bm{r})=\pm\Psi_0\neq 0$ deep 
within either of them. For microphase
separation we expect $\Psi(\bm {k})\neq 0$ for the nonzero $\bm{k}$'s
characterising, {\it e.g.\/}, a striped state.

The cross-links are taken to connect randomly chosen pairs of monomers,
without regard to their type, except for the intrinsic charge correlations
present in the un-cross-linked blend. Hence, we use the Deam-Edwards
distribution~\cite{Deam+Edwards1975} to average over the quenched
connectivity.  As a function of the amount of cross-linking, the system undergoes a gelation transition from a viscous fluid to an amorphous solid at
$\mu_{\textrm{c}}=1$~\cite{GoldbartAdvPhy1996}.  The order parameter
for the gelation transition comprises the ($g\ge 2$) moments of
the local density fluctuations:
\begin{equation}
\Omega(\bm{k}_1,\bm{k}_2, \dots ,\bm{k}_g) =\frac{1}{N}
  \sum_{a=A,B}\sum_{i=1}^{N/2}\int_0^1\!ds\,
  \Bigl[
  \big\langle e^{i\bm{k}_1\cdot\bm{r}_i^a(s)}\big\rangle
  \dots
  \big\langle e^{i\bm{k}_g\cdot\bm{r}_i^a(s)}\big\rangle
  \Bigr]\ ,
\end{equation}
and these acquire nonzero values in the gelled state, provided that
$\bm{k}_1+\bm{k}_2 \dots +\bm{k}_g=\bm{0}$. This condition is a consequence of
the macroscopic translational invariance (MTI) of the emerging amorphous
solid, which of course enforces the MTI of the density itself:
$\Omega(\bm{k}_1)=0$. In our model this homogeneity is ensured by a
sufficiently strong excluded-volume interaction. The thermal average is
denoted by $\langle\cdots\rangle$, and the quenched average over the
cross-links by $[\cdots]$.


\section{Landau free energy}

Averaging over the quenched random cross-links with help of the
replica technique gives rise to a Landau free energy
$f=\lim_{n \to 0}f_n$ in terms of the replicated order parameters:
\begin{multline}
  \label{eq:Landau}
  2n f_n(\{\Psi,\Omega\}) =
  \olsum{\hat k} \Big( \tfrac{1}{\mu}
  - g_{\textrm{D}}\big(|\hat k|\big)\!\Big)\,
  \big| \Omega\big({\hat k}\big) \big|^2
  -\,\tfrac{1}{3}\!\!\olsum{{\hat k}_{1},{\hat k}_{2},{\hat k}_{3}}
  \delta_{{\hat k}_{1}+{\hat k}_{2}+{\hat k}_{3},{\hat 0}}\,
  \Omega({\hat k}_{1})\,\Omega({\hat k}_{2})\,\Omega({\hat k}_{3}) \\
  +\sum_{\alpha=0}^n \plsum{\bm{k}}
  \Big( \tfrac{1}{\chi^{\alpha}}
  - g_{\textrm{D}}\big(|\bm{k}|\big) \Big)\,
  \big|  \Psi^\alpha(\bm{k}) \big|^2
  - \sum_{\alpha\neq\beta}
  \plsum{\bm{k}_1,\bm{k}_2}
  \Psi^{\alpha}(\bm{k}_1)\,
  \Psi^{\beta}(\bm{k}_2)\,
  \Omega(-\bm{k}_1 \hat e^{\alpha}-\bm{k}_2 \hat e^{\beta})
  +\ \cdots
\end{multline}
To simplify the notation we have introduced $(n+1)$-times replicated
wave-vectors $\hat k = (\bm{k}_0,\bm{k}_1,\ldots,\bm{k}_n)=
\sum_{\alpha=0}^n \bm{k}_{\alpha}\,\hat e^{\alpha}$,
which can be expanded in a complete orthonormal basis
$\{\hat e^{\alpha}\}_{\alpha=0}^n$ in replica space.
The symbols $\psum{{\bm k}}$ and $\osum{\hat k}$ denote
summations excluding the terms $\bm{k}=\bm{0}$ and
$\hat k=\bm{k}\,{\hat e}^{\alpha}$.
The Debye function is given by $g_{\textrm{D}}(k)\equiv
2(e^{-k^2}-1+k^2)/k^4$, and all wave-numbers have been made dimensionless with
the radius of gyration $R_{\textrm{g}}^2=La^2/2d$ of a single chain, with $d$
being the dimensionality of space.
In general, the homopolymer blend is cross-linked at one temperature,
$\chiprep^{-1}$, and measurements are performed at a distinct temperature,
$\chimeas^{-1}$. In the replica formalism this is encoded in
$\chi^{\alpha}=(\chiprep,\chimeas,\ldots,\chimeas)$.

As a first step, we apply the {\it saddle-point\/} approximation
to Eq.~(\ref{eq:Landau}) in the range of parameters
$\chiprep<1$, $\chimeas<1$ and $\mu>1$, so that $\bar\Psi=0$,
whereas the order parameter for the
gel~\cite{Castillo+Goldbart+Zippelius1994,
  Peng+Castillo+Goldbart+Zippelius1998} acquires a nonzero value: 
\begin{equation} \label{eq:Omega-saddlepoint}
  \bar\Omega(\hat{\bm{k}}) =
  \delta_{\tilde{\bm{k}},\bm{0}} \, Q \int_0^\infty\!d\tau\,
  \pi(\tau)\exp(-\hat k^2\lenloc^2/2\tau)
  \,\approx\, \delta_{\tilde{\bm{k}},\bm{0}} \cdot
  \lenloc^{-2}\;\scalf\Big(|\hat{k}|\lenloc/\sqrt{2}\Big)\,,
\end{equation}
A nonzero fraction $Q$ of particles is localised with a broad distribution of
localisation lengths, expressed in terms of a scaling function $\pi(\tau)$.
The typical localisation length $\lenloc$ is given by the distance
$\lenloc^{-2}=\mu-1$ from the gelation transition. Macroscopic translational
invariance is guaranteed by the Kronecker-$\delta$ factor, enforcing
$\tilde{\bm{k}} \equiv \sum_\alpha \bm{k}_\alpha = \bm{0}$.
The scaling function $\scalf(x)$ is related to that defined
in~\cite{GoldbartAdvPhy1996} via $\scalf(x)\equiv 2\cdot \omega\big(
\sqrt{4/3}\cdot x\big)$.


\section{Stability of the homogeneously mixed state}

Next we consider charge fluctuations and study stability
with respect to phase separation with help of the Hessian matrix
\begin{equation}
  A_{\alpha,\beta}(k)\equiv
  \delta_{\alpha,\beta}
  \big(\chi_{\alpha}^{-1}- g_{\textrm{D}}(k)\big)
  \ -\ (1-\delta_{\alpha,\beta})
  \,\lenloc^{-2}\; \scalf\big(k\lenloc\big)\,,
\end{equation}
obtained by expanding Eq.~(\ref{eq:Landau}) about the saddle point.  In the
limit $n\to 0$, the eigenvalues of $A$ are given by
$\lambda_1(k)=\chiprep^{-1}-g_{\textrm{D}}(k)$ and
$\lambda_2(k)=\chimeas^{-1}-g_{\textrm{D}}(k)+ \lenloc^{-2}\;
\scalf\big(k\lenloc\big)$. The first eigenvalue, $\lambda_1$, is not
degenerate and is always positive, as we assume that cross-linking takes place
in the homogeneously mixed state.  The second eigenvalue, $\lambda_2$, is
$n$-fold degenerate and vanishes at $\chicrit=1+1.38\, \lenloc^{-2}$ at the
nonzero wave-number $k^2_{\textrm{crit}}=1.78\,\lenloc^{-2}$. (These results
were derived from the Landau expansion, and hence are systematic only to
lowest order in~$\lenloc^{-2}$.)  We conclude that the range of temperatures
$\chimeas<\chicrit$ in which the gel is stable with respect to phase
separation is enlarged, compared to the un-cross-linked polymer blend, as one
would expect, because cross-links hinder the demixing process. In this
temperature range there are, however, frozen-in charge fluctuations, which we
shall discuss next. Later we shall discuss the instability with respect to MPS
that occurs when $\chimeas$ approaches $\chicrit$ from below.


\section{Pseudo phase diagram}

A \lqq phase diagram\rqq\ in the $\chi$-$\mu$-plane is shown in
Fig.~\ref{fig:Phase-diagram}.  The line $\mu=1$ (dashed) separates the
gel and the liquid state, which is divided into a mixed phase and a
macroscopically phase separated liquid by $\chi=1$ (solid line).  We
restrict the discussion to gels prepared from a homogeneous liquid
($\chiprep<1$) by {\it instantaneous\/} cross-linking (vertical jump);
after preparation, only the incompatibility~$\chimeas$ may be varied
(horizontal moves). The line $\chimeas=\chicrit$ (dotted) separates
the mixed gel from the microphase separated one.  The history of the
gelation process is indicated by a {\it path\/} in the diagram. Of
particular interest are three paths: {\it A\/} refers to a gel
prepared close to macroscopic phase separation, {\it B\/} to a weak
gel, and {\it C\/} to a gel close to microphase separation. Since the
properties of the gel depend on the preparation,
Fig.~\ref{fig:Phase-diagram} is not an equilibrium phase diagram.


\section{Charge glass}

Within the Gaussian expansion around the saddle point, charge fluctuations are
determined by the Hessian according to
$\langle\Psi^{\alpha}(\bm{k})\Psi^{\beta}(-\bm{k})\rangle=
(A^{-1}(k))_{\alpha,\beta}$. In particular, glassy correlations are detected
by a non-zero off-diagonal propagator:
\begin{equation}
  \label{eq:Glass}
  S_{\textrm{gl}}(k) = [\langle\Psi(\bm{k})\rangle
  \langle\Psi(-\bm{k})\rangle]
  = 
  \frac{
    \lenloc^{-2}\scalf(k\lenloc)\,\big(\chiprep^{-1}-g_{\textrm{D}}(k)
    +\lenloc^{-2}\scalf(k\lenloc)\big)}
  {(\chiprep^{-1}-g_{\textrm{D}}(k))\,\lambda_2^2(k)}\,.
\end{equation}
To discuss these fluctuations, one should note that there are three competing
length-scales: The typical localisation length $\lenloc$ of the gel, determined
by the distance to the gelation threshold; the correlation length $\lenprep$
of the preparation ensemble, given by the distance $\lenprep^{-2}=1-\chiprep$
to the demixing transition {\it before\/} cross-linking; and $\lenmeas$,
corresponding to the distance $\lenmeas^{-2}=\chicrit-\chimeas$ to the
microphase separation line.

\begin{figure}
  \vspace{-1ex} 
  \begin{minipage}[t]{0.45\textwidth}
    \center
    \includegraphics[scale=0.28]{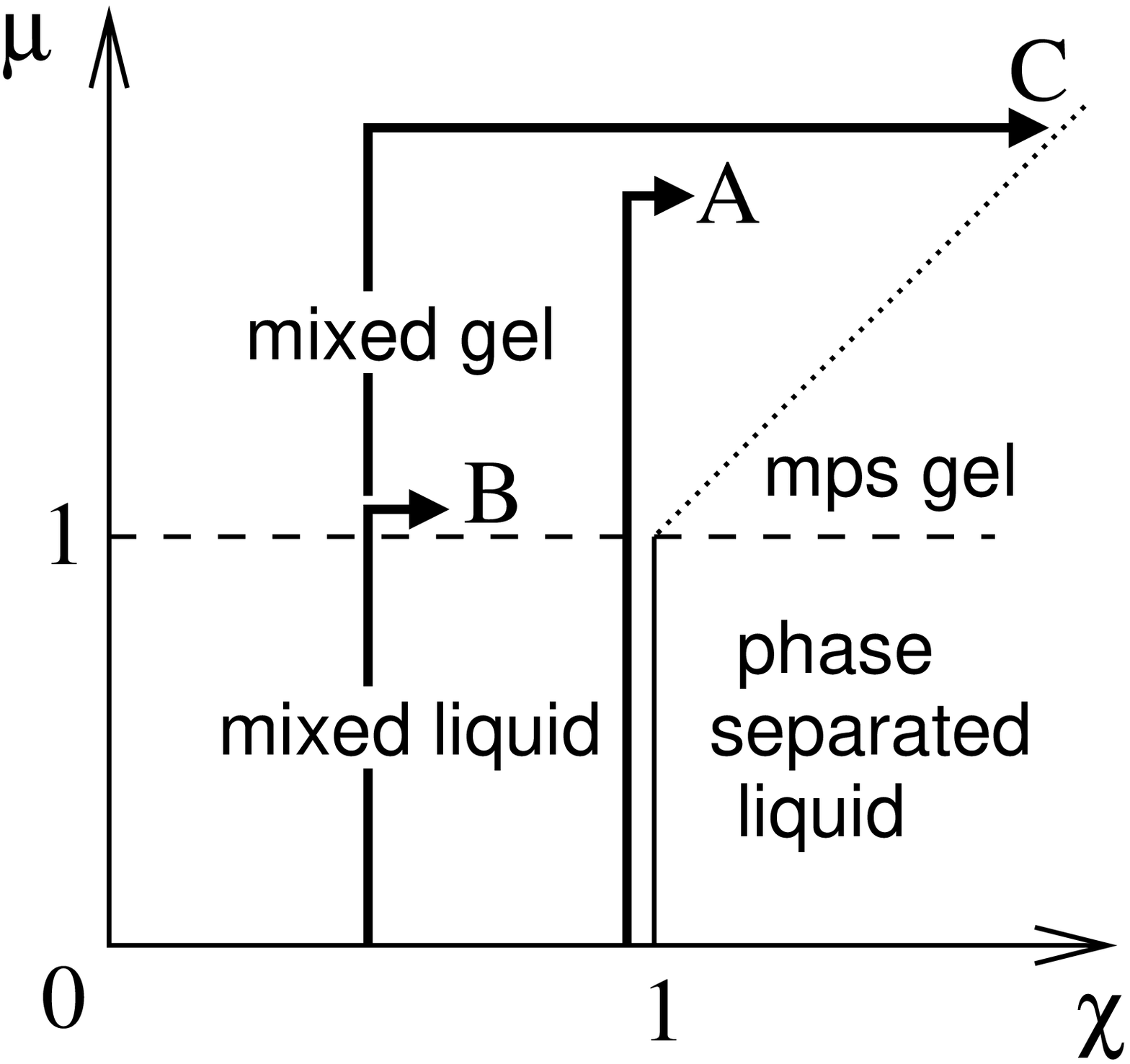}
  
    \caption{Pseudo phase diagram of the polymer blend in the
      $\chi$-$\mu$-plane. The state of the system is, however,
      history-dependent (see text for details).
      \label{fig:Phase-diagram} }
  \end{minipage}
  \hfill
  \begin{minipage}[t]{0.51\textwidth}
    \center
    \psfrag{Pksq}{\raisebox{0.5ex}{\large$k^2$}}
    \psfrag{PSrel}{\large\mbox{}\hspace{-4ex}$S_{\textrm{gl}}(k)/S_{\textrm{gl}}(0)$}
    \psfrag{1e-06}{\large\raisebox{-0.25ex}{$10^{-6}$}}
    \psfrag{1e-04}{\large\raisebox{-0.25ex}{$10^{-4}$}}
    \psfrag{1e-02}{\large\raisebox{-0.25ex}{$10^{-2}$}}
    \psfrag{1e+00}{\large\raisebox{-0.25ex}{$10^{0}$}}
    \psfrag{0.0}{\large$0.0$}
    \psfrag{0.2}{\large$0.2$}
    \psfrag{0.4}{\large$0.4$}
    \psfrag{0.6}{\large$0.6$}
    \psfrag{0.8}{\large$0.8$}
    \psfrag{1.0}{\large$1.0$}
    
    \psfrag{Pdlc}{\raisebox{0.25ex}{${}\!\lenprep^{-2}$}}
    \psfrag{-6P}{\footnotesize $\!10^{\mbox{-}6}$}
    \psfrag{-4P}{\footnotesize $\!10^{\mbox{-}4}$}
    \psfrag{-2P}{\footnotesize $\!10^{\mbox{-}2}$}
    \psfrag{0P}{\footnotesize $\!\!\!10^{0}$}
    \psfrag{-6}{\footnotesize $\!\!\!\!10^{\mbox{\,-}6}$}
    \psfrag{-4}{\footnotesize $\!\!\!\!10^{\mbox{\,-}4}$}
    \psfrag{-2}{\footnotesize $\!\!\!\!10^{\mbox{\,-}2}$}
    
    \scalebox{0.666667}{\includegraphics[width=1.3725\textwidth]
      {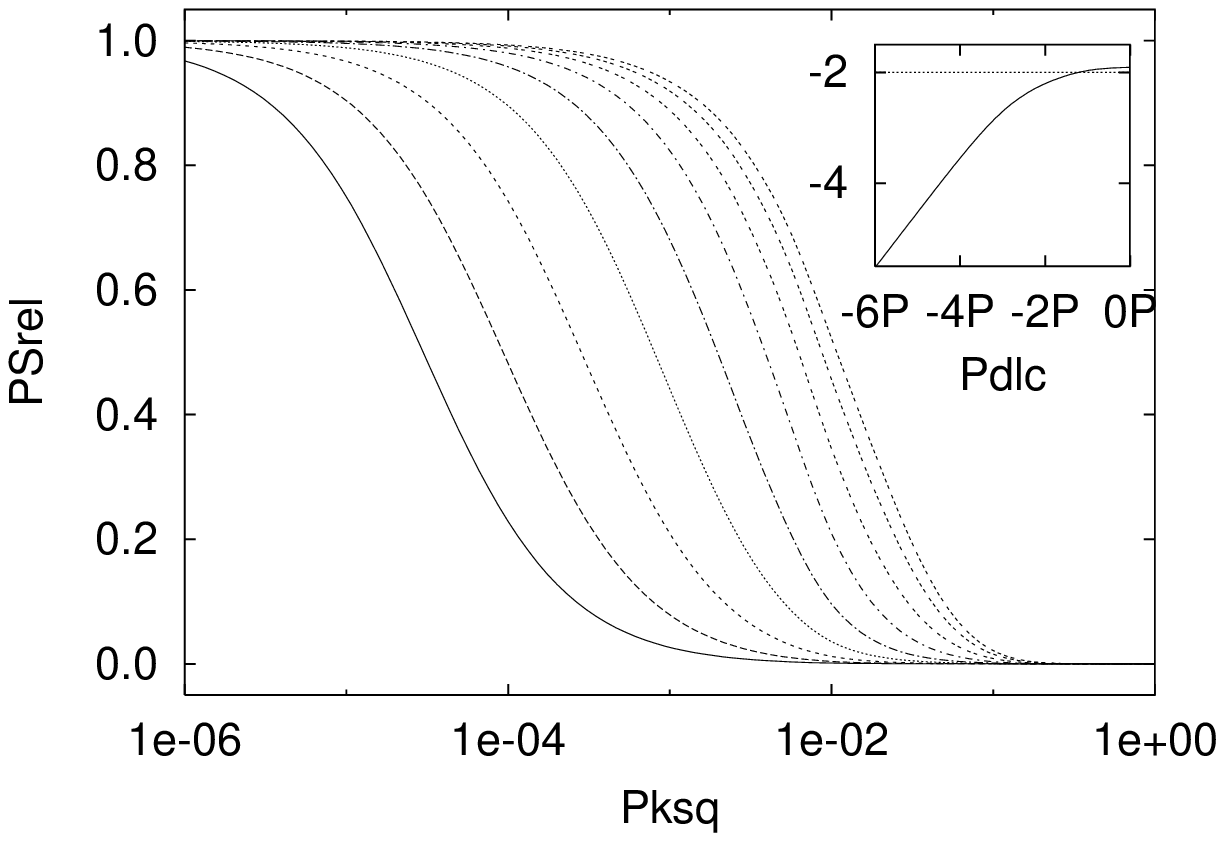}}

    \caption{Disorder-averaged $S_{\textrm{gl}}(k)/S_{\textrm{gl}}(0)$
      vs.~$k^2$ 
      for $\chimeas=0.1$, $\lenloc^2=10^2$, and
      $\lenprep^2=10^{5},10^{4.5},\ldots,10^{1}$.
      Inset: Half-width, crossing over from $\lenprep^{-2}$ to
      $\lenloc^{-2}$ (horizontal line).
      \label{fig:Sgl-crossover} }
  \end{minipage}
 
  \vspace{-3ex}
\end{figure}

If $\lenprep\gg(\lenloc,\lenmeas)$, the gel is prepared close to macroscopic
phase separation, corresponding to path {\it A\/} in
Fig.~\ref{fig:Phase-diagram}.  The thermal charge fluctuations present before
cross-linking are very long-ranged, and are almost completely frozen in by the
network. In this regime, Eq.~(\ref{eq:Glass}) reduces to $S_{\textrm{gl}}(k)
\sim (\lenprep^{-2}+k^2/3)^{-1}$, and the glassy correlations decay on the
scale $k\sim\lenprep^{-1}$, given by the fluctuations of the {\it preparation
  ensemble\/}.  

If $\lenloc\gg(\lenprep,\lenmeas)$, {\it i.e.\/}~in a weak gel (path
{\it B\/}), the network is rather wide-meshed, which limits the
preservation of fluctuations to the inverse localisation length
$\lenloc^{-1}$. In this limit, the glassy correlations are
approximately given by $ S_{\textrm{gl}}(k) \sim
\scalf(k\lenloc)/\lenloc^2$, and decay on the scale
$k\sim\lenloc^{-1}$.

The cross-over between these scales is demonstrated in
Fig.~\ref{fig:Sgl-crossover}, which shows $S_{\textrm{gl}}(k)/ S_{\textrm{gl}}
(0)$ far from MPS for $\lenloc=10$.  For the leftmost curve $\lenprep^2=10^5
\gg \lenloc^2=100$, and hence the decay occurs at $k \sim \lenprep^{-1}$. Upon
decreasing $\lenprep$ the curves are shifted to the right, until for $\lenloc
\gg \lenprep$ the decay is determined by $\lenloc$.  The inset shows the half
width at half maximum as a function of $\lenprep^{-1}$.

Close to microphase separation (path {\it C\/}), the glassy correlations are
dominated by critical fluctuations on the scale $k\approx k_{\textrm{crit}}
\sim\lenloc^{-1}$ and ultimately diverge at microphase separation.


\section{Scattering function}

The scattering intensity is determined by the diagonal correlations
\begin{equation} \label{eq:Scatter}
  S(k)+\chimeas
  =[\langle\Psi(\bm{k}) \Psi(-\bm{k})\rangle]
  = \lambda_2^{-1}(k)+ \frac{\lenloc^{-2}\scalf(k\lenloc) \,
    \big(\chiprep^{-1}-g_{\textrm{D}}(k) +\lenloc^{-2}\scalf(k\lenloc)\big)}
  {\lambda_2^2(k)\,(\chiprep^{-1}-g_{\textrm{D}}(k))}\,.
\end{equation}
We expect two main contributions to $S(k)$: The fluctuations that are present
at cross-linking and frozen in in the gel, are characterised by small
wave-numbers, being the precursor of a homogeneous phase separation. The
fluctuations that signal MPS, on the other hand, are dominant at
$k\sim\lenloc^{-1}$.

If $\lenprep\gg(\lenloc,\lenmeas)$, {\it i.e.\/} a gel prepared close
to phase separation (path {\it A\/}), the network preserves the strong
fluctuations that were present before gelation, giving a large
contribution at small wave vectors up to $k\sim\lenprep^{-1}$.

If $\lenmeas\gg(\lenloc,\lenprep)$, {\it i.e.\/} measurement close to MPS
(path {\it C\/}), the fluctuations towards microphase separation grow large,
giving rise to a peak in the scattering function at $k\approx k_\textrm{crit}
\sim\lenloc^{-1}$. At the critical point of MPS, the scattering function
diverges like $S(k) \sim \lambda_2^{-2}(k)$.

\begin{figure}
  \vspace{-2ex}
  \center
  \subfigure[$\lenloc^2=10^{1}$, $\lenmeas^2=10^{3}$, several values of
    $\lenprep^2$ (see legend).]{    \label{fig:S-crossover1a}
    \psfrag{PX0.0}{\large\raisebox{-1ex}{\large $0.0$}}
    \psfrag{PX0.1}{\large\raisebox{-1ex}{\large $0.1$}}
    \psfrag{PX0.2}{\large\raisebox{-1ex}{\large $0.2$}}
    \psfrag{PX0.3}{\large\raisebox{-1ex}{\large $0.3$}}
    \psfrag{PX0.4}{\large\raisebox{-1ex}{\large $0.4$}}
    \psfrag{PFkqu}{\raisebox{-3ex}{\Large $k^2$}}

    \psfrag{PY1e+01}{\large\raisebox{-0.5ex}{$\!\,10^{1}$}}
    \psfrag{PY1e+02}{\large\raisebox{-0.5ex}{$\!\,10^{2}$}}
    \psfrag{PY1e+03}{\large\raisebox{-0.5ex}{$\!\,10^{3}$}}
    \psfrag{PY1e+04}{\large\raisebox{-0.5ex}{$\!\,10^{4}$}}
    \psfrag{PY1e+05}{\large\raisebox{-0.5ex}{$\!\,10^{5}$}}
    \psfrag{PY1e+06}{\large\raisebox{-0.5ex}{$\!\,10^{6}$}}
    \psfrag{PFS(k)}{\Large\raisebox{.5ex}{\large $S(k)$}}

    \psfrag{PFc01e-1}{\footnotesize $\hspace{-7.5ex}\lenprep^{-2}=10^{-1}$}
    \psfrag{PFc01e-3}{\footnotesize $\hspace{-7.5ex}\lenprep^{-2}=10^{-3}$}
    \psfrag{PFc01e-5}{\footnotesize $\hspace{-7.5ex}\lenprep^{-2}=10^{-5}$}

    \scalebox{.66667}{\includegraphics[width=.7\textwidth]{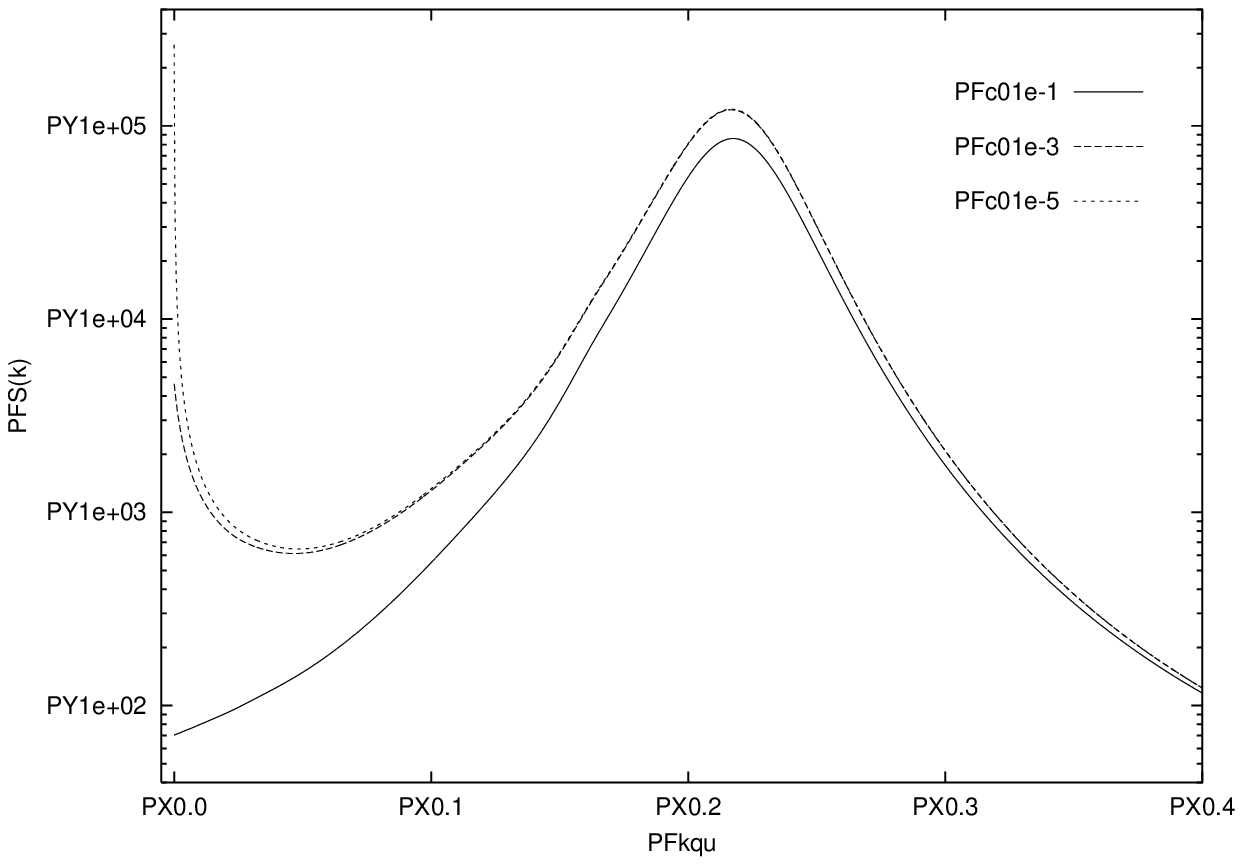}}
  }
  \hfill
  \subfigure[$\lenloc^2=10^{1}$, $\lenprep^2=10^{3}$, several values of
    $\lenmeas^2$ (see legend).]{    \label{fig:S-crossover1b}
    \psfrag{PX0.0}{\large\raisebox{-1ex}{\large $0.0$}}
    \psfrag{PX0.1}{\large\raisebox{-1ex}{\large $0.1$}}
    \psfrag{PX0.2}{\large\raisebox{-1ex}{\large $0.2$}}
    \psfrag{PX0.3}{\large\raisebox{-1ex}{\large $0.3$}}
    \psfrag{PX0.4}{\large\raisebox{-1ex}{\large $0.4$}}
    \psfrag{PFkqu}{\raisebox{-3ex}{\Large $k^2$}}

    \psfrag{PY1e+00}{\large\raisebox{-0.5ex}{$\!\,10^{0}$}}
    \psfrag{PY1e+02}{\large\raisebox{-0.5ex}{$\!\,10^{2}$}}
    \psfrag{PY1e+04}{\large\raisebox{-0.5ex}{$\!\,10^{4}$}}
    \psfrag{PY1e+06}{\large\raisebox{-0.5ex}{$\!\,10^{6}$}}
    \psfrag{PY1e+08}{\large\raisebox{-0.5ex}{$\!\,10^{8}$}}
    \psfrag{PFS(k)}{\Large\raisebox{.5ex}{\large $S(k)$}}

    \psfrag{PFcr1e-1}{\footnotesize $\hspace{-8ex}\lenmeas^{-2}=10^{-1}$}
    \psfrag{PFcr1e-3}{\footnotesize $\hspace{-8ex}\lenmeas^{-2}=10^{-3}$}
    \psfrag{PFcr1e-5}{\footnotesize $\hspace{-8ex}\lenmeas^{-2}=10^{-5}$}

    \scalebox{.66667}{\includegraphics[width=.7\textwidth]{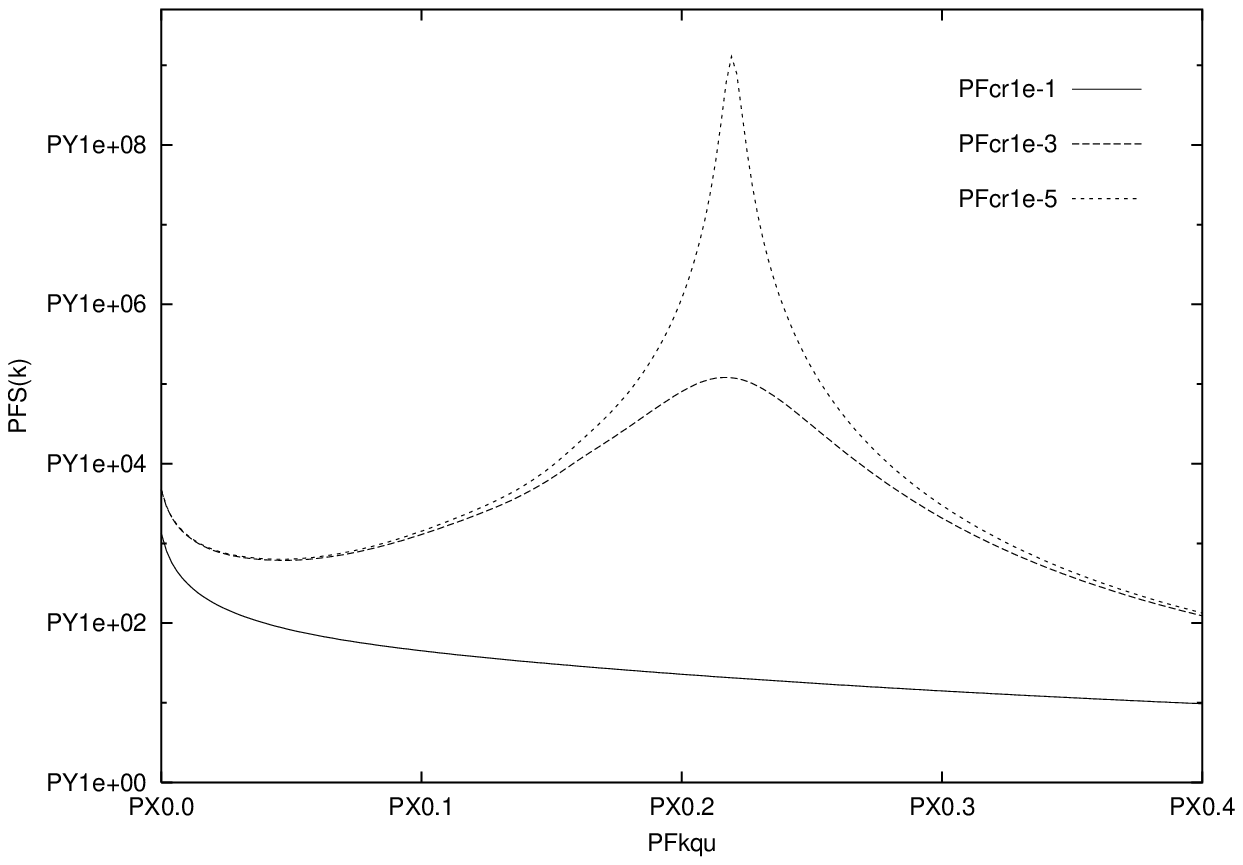}}
  }

  \vspace{-3ex}
  \caption[Scattering function]{
    Scattering function $S(k)$ vs.~$k^2$.
  }
  \label{fig:S-crossover1}
\end{figure}

The contributions of both regimes can be seen in Fig.~\ref{fig:S-crossover1}.
Panel~(a) shows $S(k)$ for a fixed distance $\lenmeas^{-2}$ from the MPS
transition with $\lenprep^{-2}$ decreasing towards (pre-cross-linking) phase
separation. We observe increasingly stronger weight to $S(k)$ at small
wave-numbers.  Panel~(b) shows $S(k)$ at a fixed distance $\lenprep^{-2}$ from
phase separation before cross-linking. Upon decreasing the distance
$\lenmeas^{-2}$ to MPS, the scattering function $S(k)$ is dominated more and
more by fluctuations at $k_{\textrm{crit}}$.

\begin{figure}
  \vspace{-2ex}
  \center
  \subfigure[$\lenloc^2=10^{3}$, $\lenmeas^2=10^{1}$, several values of
  $\lenprep^2$ (see legend).
  Vertical line: $k^2=3\,\lenmeas^{-2}$]
  {      \label{fig:S-crossover2b}
    \psfrag{PX1e-09}{\large\raisebox{-1ex}{\large $10^{-9}$}}
    \psfrag{PX1e-06}{\large\raisebox{-1ex}{\large $10^{-6}$}}
    \psfrag{PX1e-03}{\large\raisebox{-1ex}{\large $10^{-3}$}}
    \psfrag{PX1e+00}{\large\raisebox{-1ex}{\large $10^{0}$}}
    \psfrag{PFkqu}{\raisebox{-1.5ex}{\large $k^2$}}
    \psfrag{PY1e+00}{\large\raisebox{-0.5ex}{$\!\,10^{0}$}}
    \psfrag{PY1e+01}{}
    \psfrag{PY1e+02}{\large\raisebox{-0.5ex}{$\!\,10^{2}$}}
    \psfrag{PY1e+03}{}
    \psfrag{PY1e+04}{\large\raisebox{-0.5ex}{$\!\,10^{4}$}}
    \psfrag{PFS(k)}{\large\raisebox{.5ex}{\large $S(k)$}}

    \psfrag{PFc01e-2}{\footnotesize $\hspace{-7.5ex}\lenprep^{-2}=10^{-2}$}
    \psfrag{PFc01e-4}{\footnotesize $\hspace{-7.5ex}\lenprep^{-2}=10^{-4}$}
    \psfrag{PFc01e-6}{\footnotesize $\hspace{-7.5ex}\lenprep^{-2}=10^{-6}$}
    \psfrag{PFc01e-8}{\footnotesize $\hspace{-7.5ex}\lenprep^{-2}=10^{-8}$}

    \scalebox{.66667}{\includegraphics[width=.7\textwidth]{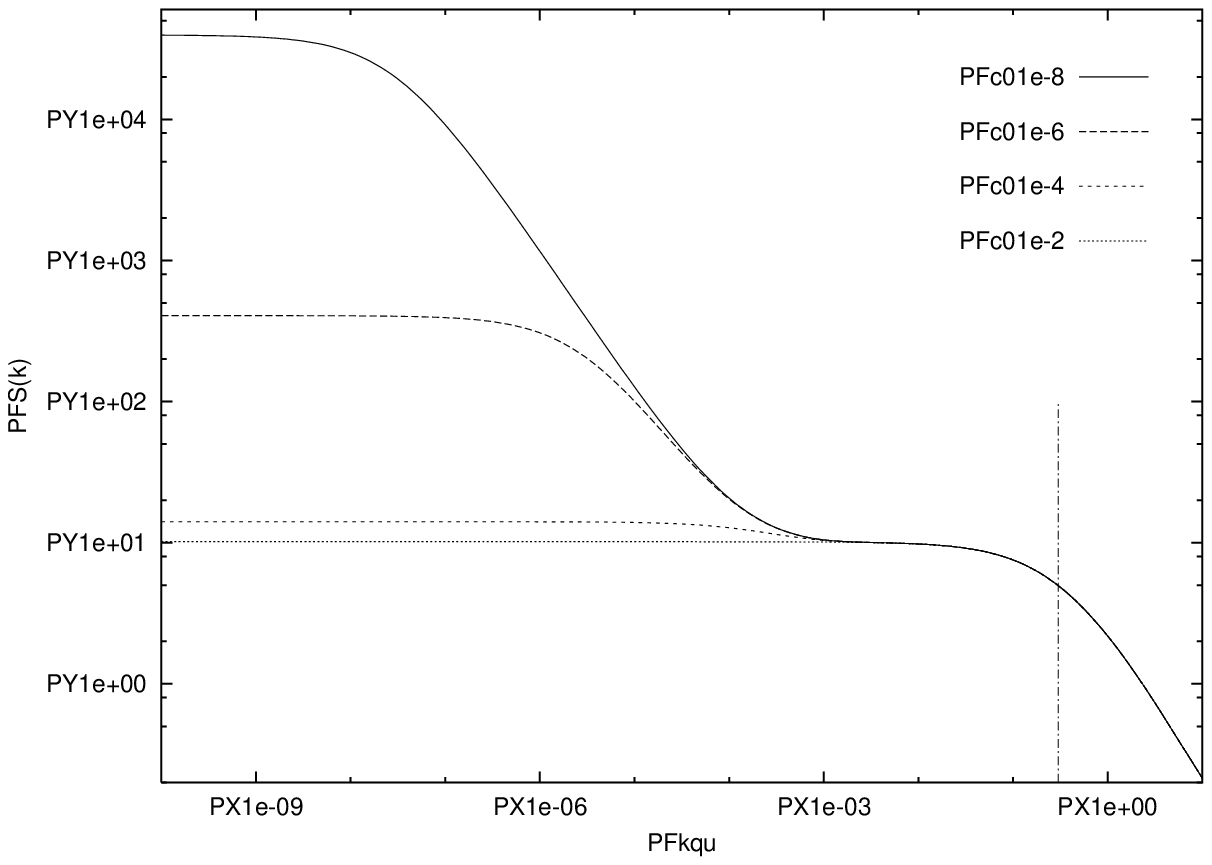}}
  }
  \hfill
  \subfigure[$\lenloc^2=10^{2}$, $\lenprep^2=10^{1}$, several values of
  $\lenmeas^2$ (see legend).]{       \label{fig:S-crossover2a}
    \psfrag{PX1e-04}{\large\raisebox{-1ex}{\large $10^{-4}$}}
    \psfrag{PX1e-03}{\large\raisebox{-1ex}{\large $10^{-3}$}}
    \psfrag{PX1e-02}{\large\raisebox{-1ex}{\large $10^{-2}$}}
    \psfrag{PX1e-01}{\large\raisebox{-1ex}{\large $10^{-1}$}}
    \psfrag{PX1e+00}{\large\raisebox{-1ex}{\large $10^{0}$}}
    \psfrag{PX1e+01}{\large\raisebox{-1ex}{\large $10^{1}$}}
    \psfrag{PFkqu}{\raisebox{-1.5ex}{\large $k^2$}}
    \psfrag{PY1e+00}{\large\raisebox{-0.5ex}{$\!\,10^{0}$}}
    \psfrag{PY1e+01}{}
    \psfrag{PY1e+02}{\large\raisebox{-0.5ex}{$\!\,10^{2}$}}
    \psfrag{PY1e+03}{}
    \psfrag{PY1e+04}{\large\raisebox{-0.5ex}{$\!\,10^{4}$}}
    \psfrag{PY1e+05}{}
    \psfrag{PY1e+06}{\large\raisebox{-0.5ex}{$\!\,10^{6}$}}
    \psfrag{PFS(k)}{\large\raisebox{.5ex}{\large $S(k)$}}

    \psfrag{PFcr1e-1}{\footnotesize $\hspace{-8ex}\lenmeas^{-2}=10^{-1}$}
    \psfrag{PFcr1e-2}{\footnotesize $\hspace{-8ex}\lenmeas^{-2}=10^{-2}$}
    \psfrag{PFcr1e-3}{\footnotesize $\hspace{-8ex}\lenmeas^{-2}=10^{-3}$}
    \psfrag{PFcr1e-4}{\footnotesize $\hspace{-8ex}\lenmeas^{-2}=10^{-4}$}

    \scalebox{.66667}{\includegraphics[width=.7\textwidth]{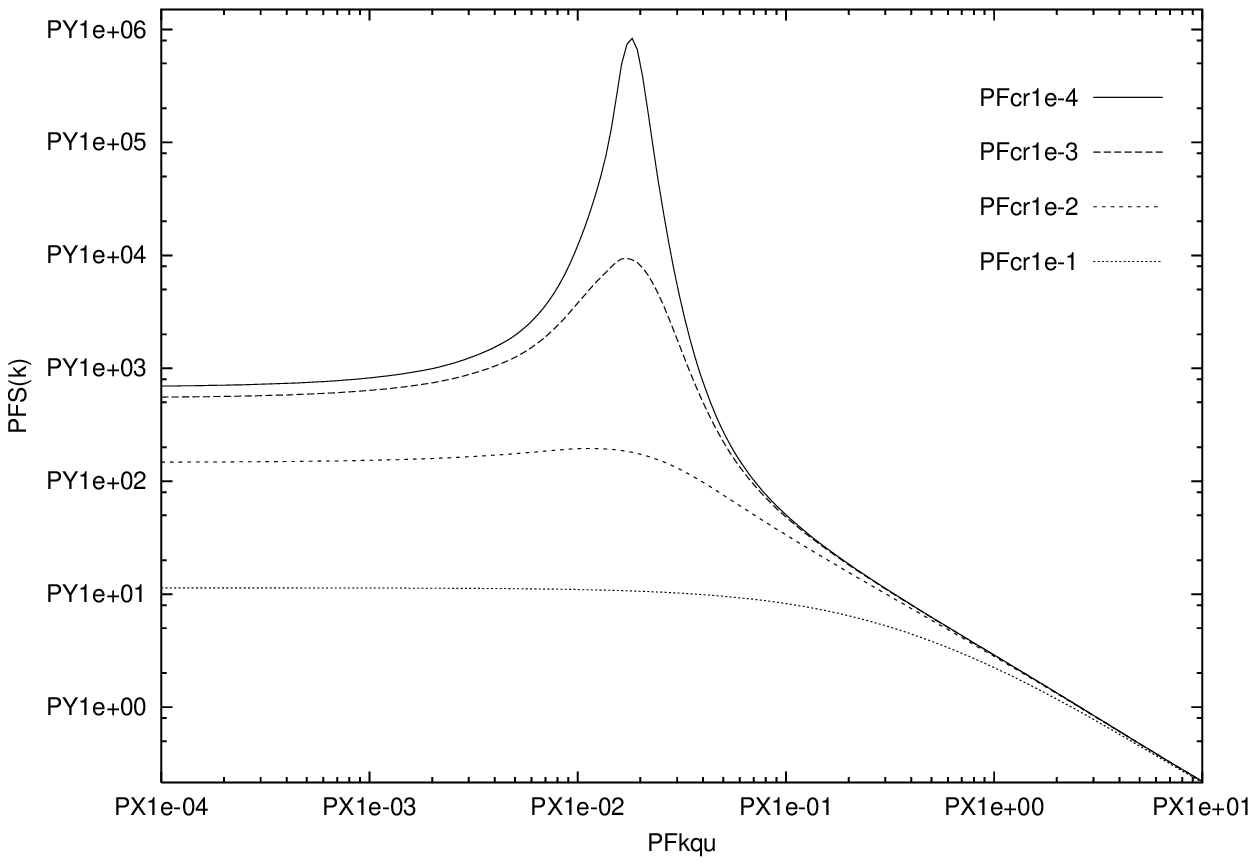}}
  }

\vspace{-3ex}
\caption[Scattering function]{
    Scattering function $S(k)$ vs.~$k^2$.
  }
  \label{fig:S-crossover2}

\end{figure}

If $\lenloc\gg(\lenprep,\lenmeas)$, {\it i.e.\/}~in a weak gel (path {\it
  B\/}), the frozen-in fluctuations are small on all length-scales.  Hence,
the scattering function is dominated by purely thermal fluctuations, seemingly
towards macroscopic demixing, because the gel is too weak to severely restrict
phase separation on the length scale $\lenmeas$. In this limit,
Eq.~(\ref{eq:Scatter}) reduces to $ S(k) = \lenmeas^{-2}+k^2/3$, which decays
on a scale $k\sim\lenmeas^{-1}$ and is shown as the lowest curve in
Fig.~\ref{fig:S-crossover2}.  Decreasing the distance to the (pre-gelation)
phase separation point, $\lenprep^{-2}$, we observe the build-up of intensity
at small wave-numbers (panel a). Decreasing the distance to microphase
separation, $\lenmeas^{-2}$, leads to an increasingly stronger peak at
$k_{\textrm{crit}}^2\approx1.78\,\lenloc^{-2}$ (panel b).

\section{Microphase separation}

The analysis of the microphase separated state requires supplementing the
Landau expansion~(\ref{eq:Landau}) by fourth-order terms. To minimise this
free energy variationally, we consider a simple, replica-symmetric, lamellar
microphase state:
\begin{equation} \label{eq:mps-ansatz}
  \Psi^{\alpha}_{\bm{k}} = \begin{cases}
    \, 0, & \textrm{for }\alpha=0, \\
    \,\tfrac{1}{\sqrt{2}} \bigl( \delta_{\bm{k},\bm{q}}
    + \delta_{\bm{k}, -\bm{q}} \bigr)\, \bar{\Psi},
    & \textrm{otherwise}\,.
  \end{cases}
\end{equation}
Note that, because the melt is cross-linked in the homogeneously mixed state
($\chiprep<1$), the zeroth replica is treated separately. Inserting the
Ansatz~(\ref{eq:mps-ansatz}) into the fourth-order expansion of the free
energy we obtain
\begin{equation}
  \lim_{n\rightarrow0}
  f_n\big(\bar{\Psi},\,q\big) =
  2\lambda_2(q)\,\bar{\Psi}^2 + g_4(q) \, \bar{\Psi}^4\,,
\end{equation}
where~$g_4(q)=1-\tfrac{2}{3}q^2+\mathcal{O}(q^4)$ denotes a Debye-like vertex
function. Minimisation with respect to both the amplitude $\bar \Psi$ and the
wave-number $q$ yields 
\begin{equation}
  \bar \Psi^2 \sim \chimeas-\chicrit \qquad \mbox{and} \qquad
  q_{\rm opt}^2-k_{\rm crit}^2\sim \chimeas-\chicrit\,.
\end{equation}
We presume that the free energy would be lowered further in a state that is not
characterised simply by a single wave-vector, as above; instead the direction
of the wave vector should vary locally to adjust to the random static charge
fluctuations induced by the cross-links.


\acknowledgments

We thank Xiangjun Xing for enlightening discussions.
This work was supported in part by
the Deutsche Forschungsgemeinschaft through
SFB~602 (AZ, CW),
Grant No.~Zi~209/6-1 (AZ), and GRK~782 (CW),
and by the U.S.~National Science Foundation
through grant NSF DMR02-05858 (PMG).


\end{document}